\documentclass[letterpaper,twocolumn,prl,aps,superscriptaddress,amsmath,amssymb,floatfix]{revtex4-1}
\usepackage{mathptmx}
\usepackage[latin9]{inputenc}
\usepackage{color}
\usepackage{amsmath}
\usepackage{amssymb}
\usepackage{graphicx}
\usepackage{esint}
\usepackage[unicode=true,
 bookmarks=true,bookmarksnumbered=false,bookmarksopen=false,
 breaklinks=false,pdfborder={0 0 1},backref=false,colorlinks=true]
 {hyperref}
\hypersetup{
 linkcolor=magenta,urlcolor=blue,citecolor=blue,pdfstartview={FitH},hyperfootnotes=false}

\makeatletter

\usepackage{textcomp}
\usepackage{epstopdf}

\pdfpageheight\paperheight
\pdfpagewidth\paperwidth

\providecommand{\tabularnewline}{\\}

\@ifundefined{textcolor}{}{%
 \definecolor{BLACK}{gray}{0}
 \definecolor{WHITE}{gray}{1}
 \definecolor{RED}{rgb}{1,0,0}
 \definecolor{GREEN}{rgb}{0,1,0}
 \definecolor{BLUE}{rgb}{0,0,1}
 \definecolor{CYAN}{cmyk}{1,0,0,0}
 \definecolor{MAGENTA}{cmyk}{0,1,0,0}
 \definecolor{YELLOW}{cmyk}{0,0,1,0}
}

\usepackage{xcolor}\usepackage{soul}
\newcommand{\bra}[1]{\ensuremath{\left\langle#1\right|}}
\newcommand{\ket}[1]{\ensuremath{\left|#1\right\rangle}}

\definecolor{blue}{rgb}{0,0,1}
\definecolor{red}{rgb}{0,0,0}
\definecolor{green}{rgb}{0,1,0}
\newcommand{\red}[1]{\textcolor{red}{ #1}}

\usepackage{soul}

\makeatother
\begin{document}

\title{Experimental Implementation of a Qubit-Efficient Variational Quantum Eigensolver with Analog Error Mitigation on a Superconducting Quantum Processor}

\author{Yuwei~Ma}
\thanks{These authors contributed equally to this work.}
\affiliation{Center for Quantum Information, Institute for Interdisciplinary Information Sciences, Tsinghua University, Beijing 100084, China}
\author{Weiting~Wang}
\thanks{These authors contributed equally to this work.}
\affiliation{Center for Quantum Information, Institute for Interdisciplinary Information Sciences, Tsinghua University, Beijing 100084, China}
\author{Xianghao~Mu}
\affiliation{Center for Quantum Information, Institute for Interdisciplinary Information Sciences, Tsinghua University, Beijing 100084, China}
\author{Weizhou~Cai}
\affiliation{Center for Quantum Information, Institute for Interdisciplinary Information Sciences, Tsinghua University, Beijing 100084, China}
\author{Ziyue~Hua}
\affiliation{Center for Quantum Information, Institute for Interdisciplinary Information Sciences, Tsinghua University, Beijing 100084, China}
\author{Xiaoxuan~Pan}
\affiliation{Center for Quantum Information, Institute for Interdisciplinary Information Sciences, Tsinghua University, Beijing 100084, China}
\author{Dong-Ling~Deng}
\affiliation{Center for Quantum Information, Institute for Interdisciplinary Information Sciences, Tsinghua University, Beijing 100084, China}
\affiliation{Shanghai Qi Zhi Institute, 41th Floor, AI Tower, No. 701 Yunjin Road, Xuhui District, Shanghai 200232, China}
\affiliation{Hefei National Laboratory, Hefei 230088, China}
\author{Rebing Wu}
\affiliation{Center for Intelligent and Networked Systems,
Department of Automation, Tsinghua University, Beijing 100084, P. R. China}
\affiliation{Hefei National Laboratory, Hefei 230088, China}
\author{Chang-Ling~Zou}
\email{clzou321@ustc.edu.cn}
\affiliation{CAS Key Laboratory of Quantum Information, University of Science and Technology of China, Hefei, Anhui 230026, P. R. China}
\affiliation{Hefei National Laboratory, Hefei 230088, China}
\author{Lei Wang}
\email{wanglei@iphy.ac.cn}
\affiliation{Beijing National Laboratory for Condensed Matter Physics and Institute of Physics, \\Chinese Academy of Sciences, Beijing 100190, China}
\affiliation{Hefei National Laboratory, Hefei 230088, China}
\author{Luyan~Sun}
\email{luyansun@tsinghua.edu.cn}
\affiliation{Center for Quantum Information, Institute for Interdisciplinary Information Sciences, Tsinghua University, Beijing 100084, China}
\affiliation{Hefei National Laboratory, Hefei 230088, China}


\begin{abstract}
We experimentally demonstrate a qubit-efficient variational quantum eigensolver (VQE) algorithm using a superconducting quantum processor, employing minimal quantum resources with only a transmon qubit coupled to a high-coherence photonic qubit. By leveraging matrix product states to compress the quantum state representation, we simulate an $N+1$-spin circular Ising model with a transverse field. Furthermore, 
we develop an analog error mitigation approach through zero-noise extrapolation by introducing a precise noise injection technique for the transmon qubit. As a validation, we apply our error-mitigated qubit-efficient VQE in determining the ground state energies of a 4-spin Ising model. Our results demonstrate the feasibility of performing quantum algorithms with minimal quantum resources while effectively mitigating the impact of noise, offering a promising pathway to bridge the gap between theoretical advances and practical implementations on current noisy intermediate-scale quantum devices.
\end{abstract}

\maketitle


\section{Introduction}

Quantum technology has seen rapid advancement, from its theoretical beginnings~\cite{Benioff1980,Feynman1982} to the development of noisy intermediate-scale quantum (NISQ) devices~\cite{Preskill2018quantumcomputingin}. These devices, which leverage tens to hundreds of qubits, are capable of solving problems that challenge classical computers, marking an important step toward realizing quantum advantage~\cite{arute2019quantum,USTC2021_PRL}. Despite these achievements, NISQ devices face considerable limitations, including shallow circuit depths and the inability to implement full quantum error correction, which restrict the scope of quantum algorithms that can be executed effectively~\cite{TILLY20221review,Bharti_RMP2022}. 
Among the algorithms designed for these NISQ devices, the variational quantum eigensolver (VQE) stands out due to its hybrid quantum-classical approach, which has proven effective in finding the ground state energies of quantum systems~\cite{peruzzo2014a,McClean_2016,Wecker_PRA042303}. The VQE has already been used to compute molecular energies and solve quantum many-body problems~\cite{kandala2017hardware-efficient,GuoNP2024}, generating significant interest in both research and industry~\cite{Deglmann2015review,Cao_2018drug}. However, the practical implementation of VQE on NISQ devices remains challenging due to the constraints imposed by high noise level and limited qubit resources. These challenges necessitate the development of qubit-efficient strategies~\cite{Takeshita_PRX2020} and error mitigation techniques~\cite{Li_PRX2017_ERROR,Temme_EM2017PRL,kandala2019error,Colless_PRX2018,CaiRevModPhys2023} to improve the scalability and accuracy of quantum simulations.

Qubit-efficient methods, such as those leveraging matrix product states (MPS), tensor networks, and circuit-cutting techniques, reduce the number of qubits required for simulations while maintaining accuracy~\cite{LiuPhysRevResearch2019,Yuan_PRL_TN2021,Peng_PRL_2020,Ying_PRL2023_fewerQ,DeCross2023PRX,Niu2022PRXQuantum,Huggins_2019IOP}. These approaches enable the simulation of larger quantum systems on current hardware by effectively managing qubit resources, making them promising solutions for scaling quantum computations within the constraints of NISQ devices. \red{In particular, MPS provides a compact and efficient representation of quantum states, particularly for systems with limited entanglement. On a quantum computer, MPS can be implemented using fewer qubits than the physical degrees of freedom by sequentially measuring and reusing qubits. This approach effectively increases the bond dimension, allowing the representation of more entanglement with fewer qubits~\cite{schollwock2011density,Cirac2021RMP}. For example, it has been shown that even a 1D cluster state of arbitrary length can be represented using just two qubits~\cite{LiuPhysRevResearch2019}}. Additionally, error mitigation techniques, such as zero-noise extrapolation and probabilistic error cancellation, play a crucial role in reducing computational errors without the need for full error correction~\cite{Huang_SCP_2023,TILLY20221review,Temme_EM2017PRL}. These methods are particularly suited for NISQ devices, where the overhead required for full quantum error correction remains prohibitively high~\cite{Preskill2018quantumcomputingin}.


\begin{figure*}
\includegraphics{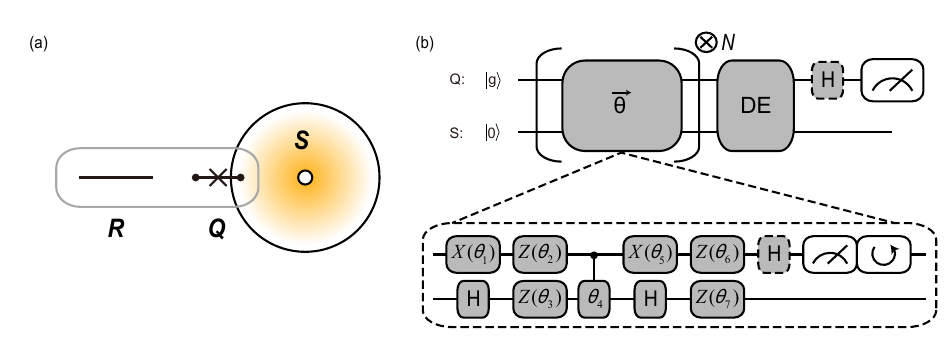}
\caption{(a) Schematic of the experimental device. The device consists of a transmon qubit ($Q$) dispersively coupled to a high-quality factor storage cavity ($S$) and a low-quality factor readout cavity ($R$). The transmon qubit serves as the reusable qubit, with real-time feedback enabling reset after each iteration. The storage cavity functions as the mediating qubit, facilitating quantum entanglement due to its longer lifetime. The readout cavity plays crucial roles in both measurement and error mitigation. (b) Circuit diagram of the VQE. We utilize one reusable qubit and one mediating qubit to execute $N$ layers of a parameterized circuit. The circuit is specifically tailored for our experimental setup, serving as the variational ansatz and efficiently representing an $N+1$-qubit quantum state as \red{an} MPS with small bond dimensions. Since measurement can only be performed on the transmon qubit, the information in the storage cavity is decoded \red{(DE)} to the transmon qubit before measurement. The Hadamard gate in the dashed-lined box is used selectively to sample the expectation values of different Pauli operators in the target Hamiltonian.
}
\label{fig:concept}
\end{figure*}

In this work, we experimentally implement a qubit-efficient VQE algorithm using a superconducting quantum processor. By leveraging MPS~\cite{LiuPhysRevResearch2019}, we reduce the qubit requirements for simulating an $N+1$-spin circular Ising model with a transverse field. Specifically, we simulate a 4-spin system using just two physical qubits, demonstrating the potential for scaling quantum simulations with fewer resources. To further enhance the accuracy of our simulations, we employ analog error mitigation using zero-noise extrapolation, which mitigates errors by extrapolating results from multiple noise levels. This combination of qubit-efficient algorithms and error mitigation allows us to push the boundaries of what is achievable on current quantum hardware, highlighting the feasibility of tackling larger quantum problems within the constraints of existing technologies.	

\section{Principle and Experimental Setup}

As shown in Fig.~\ref{fig:concept}(a), we experimentally implement a qubit-efficient VQE algorithm using a superconducting quantum processor. The quantum system consists of a transmon qubit ($Q$) dispersively coupled to a high-quality factor storage cavity ($S$)~\cite{xu2020demonstrationPRL,ma2020errorPASS,Cai2021_REVIEW,SM}. The quantum algorithm is realized by a \red{sequence} of single-qubit gates and two-qubit gates, driven by the interactions between the transmon qubit and the storage cavity. The Hamiltonian, including the driving term, can be written as:	
\begin{equation}
	H=-\chi_\mathrm{qs}a^\dagger a\sigma_\mathrm{z}/2+\Omega_\mathrm{q}(t)\sigma_++\Omega_\mathrm{q}^*(t)\sigma_-+\Omega_\mathrm{s}(t)a^\dagger+\Omega_\mathrm{s}^*(t)a,
	\label{eq:systemH}
\end{equation} 
where $\sigma_\mathrm{z},\,\sigma_+,\,\sigma_-$ denote the Pauli operators of the transmon qubit, $a$ ($a^\dagger$) is the annihilation (creation) operator acting on the photonic state in the storage cavity, $\chi_\mathrm{qs}$ represents the coefficient of the dispersive coupling, and $\Omega_\mathrm{q}^*(t)$ and $\Omega_\mathrm{s}^*(t)$ are the complex drive amplitudes applied to the transmon qubit and the storage cavity, respectively. The transmon qubit serves as the reusable qubit, with its computational basis defined by the ground state $\ket{g}$ and excited state $\ket{e}$. The storage cavity provides the photonic qubit, with its basis states defined by the vacuum state $\ket{0}_\mathrm{s}$ and single photon number state $\ket{1}_\mathrm{s}$. Given the longer lifetime of the photonic qubit compared to the transmon qubit, the photonic qubit effectively mediates quantum entanglement throughout the quantum circuit. The transmon qubit is readout via a stripline cavity, as denoted by $R$ in Fig.~\ref{fig:concept}(a), which plays a crucial role in the error mitigation techniques applied in this experiment.

Our experimental system manifests the simplest two-qubit quantum system, while allowing the implementation of general quantum algorithms through a  qubit-efficient approach. This approach leverages the reuse of the transmon qubit throughout the process of VQE algorithm~\cite{LiuPhysRevResearch2019}, drawing inspiration from tensor network representations, particularly MPS~\cite{Ostlund1995PRL,Vidal2003PRL}. By efficiently compressing quantum state information while preserving accuracy, this technique enables the simulation of larger systems with fewer qubits. In contrast to traditional VQE methods, which typically require a direct one-to-one mapping between qubits and quantum states, the qubit-efficient approach reinitializes and reuses qubits at different stages of the computation, with mediating qubits preserving the necessary entanglement between subsystems. The MPS framework limits the bond dimension, which controls the amount of entanglement captured, effectively balancing the available qubit resources with the complexity of the quantum state.

As shown in Fig.~\ref{fig:concept}(b), the qubit-efficient VQE algorithm is represented by $N$ layers of a parameterized circuit implemented on $S$ and $Q$ to generate an $N+1$-qubit state, with the detailed configuration of each layer shown in the inset. The gates in this circuit are specifically tailored to our hardware. Single-qubit gates on the transmon qubit are executed via resonant drives, with each gate parameterized by the amplitude and phase of the complex driving field $\Omega_\mathrm{q}^*(t)$. Operations on the photonic qubit, defined within the Fock state basis, are implemented with the assistance of the transmon qubit, requiring more intricate control on the composite system. Specifically, the Hadamard gate on the photonic qubit is implemented using a numerically optimized pulse, while the phase gate is performed by adjusting the phase of $\Omega_\mathrm{s}^*(t)$ through frame changes in the control software. The controlled-phase gate arises naturally from the dispersive coupling between the transmon and the photonic qubit, with the interaction time serving as a tunable control parameter. Each layer introduces seven variational parameters, and it can be proved that a single layer is sufficient to prepare an arbitrary two-qubit state from a given input state. However, due to the limited bond dimension, achieving arbitrary $N+1$-qubit states across $N$ layers is not always feasible. Nevertheless, this limitation enhances efficiency by focusing on the relevant subspace of quantum states necessary for the problem at hand.

Given the presence of noise in the quantum system, implementing an effective error mitigation strategy is crucial. Previously, zero-noise extrapolation has been employed by stretching gate time, i.e., changing the coherent interaction strengths, to scale the effective noise strengths in each quantum gate~\cite{kandala2019error}. In contrast, our approach differs by injecting noise through the direct control of the dominant decoherence rates in our system, which we refer to as analog error mitigation. This approach is widely applicable to various experimental systems, as certain coherent interaction strengths may not be adjustable in practice. Specifically, we achieve this by modifying the longitudinal relaxation time $T_1$ and the transverse relaxation time $T_2^*$ of the transmon qubit. The evolution of the quantum state ($\rho$) under the coherent Hamiltonian $H$ and \red{noises} can be described by the Lindblad master equation:
\begin{equation}
	\dot{\rho}=-i[H(t),\rho]+\lambda \mathcal{L}(\rho),
	\label{eq:TimeEvo}
\end{equation} 
where $\lambda$ is the noise strength and $\mathcal{L}$ is the Lindblad superoperator that describes the decoherence processes. By treating the noises as perturbations to the quantum evolution, the expectation value of an observable of interest $E(\lambda)$ can be expanded as a power series around its zero-noise value $E^*$ ~\cite{Li_PRX2017_ERROR,Temme_EM2017PRL,kandala2019error}:
\begin{equation}	E(\lambda)=E^*+\sum_{k=1}^{n}a_k\lambda^k+\mathcal{O}(\lambda^{n+1}),
\label{eq:Eexpand}
\end{equation} 
where $a_k$ are coefficients that depend on the specific details of the noise model and the observable being measured. By experimentally obtaining the expectation values $E(c_i\lambda)$ at different noise levels $c_i\lambda$, where $c_i$ are scaling factors, we can extrapolate back to the zero-noise value $E^*$. 


\begin{figure}
\includegraphics{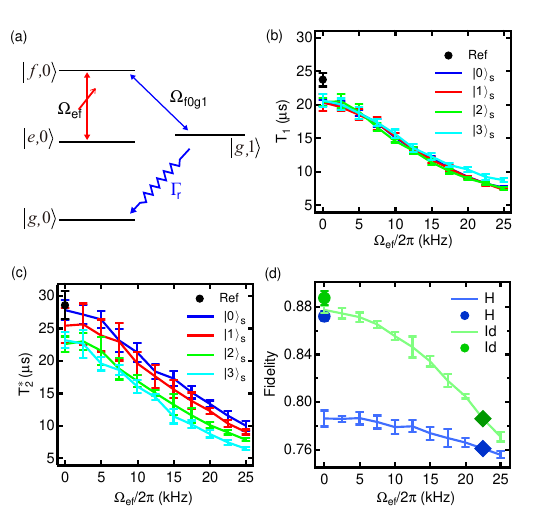}
\caption{(a) Schematic of the qubit noise injection technique, where a strong drive $\Omega_\mathrm{f0g1}$ is applied between the $\ket{f,0}_\mathrm{qr}$ and $\ket{g,1}_\mathrm{qr}$ states to control the damping rate of the transmon qubit via the low-quality factor readout cavity. (b) Longitudinal relaxation time $T_1$ and (c) transverse relaxation time $T_2^*$ of the transmon qubit as a function of the drive amplitude $\Omega_\mathrm{ef}$ for different Fock states in the storage cavity. 
(d) Gate fidelity of the Hadamard gate and identity gate as a function of $\Omega_\mathrm{ef}$. The gate fidelity is defined as the fidelity of the Pauli transfer matrix, obtained via process tomography in the experiment. The green data represent the gate fidelity when no pump is applied $(\Omega_\mathrm{f0g1}=0)$, while the blue data show the fidelity with $\Omega_\mathrm{f0g1}$ activated. The diamonds indicate the parameter $c = 2.2$, where both gate fidelity and coherence time degrade by the same factor.}
\label{fig:analogEM}
\end{figure}

\begin{figure*}[ht]
\includegraphics{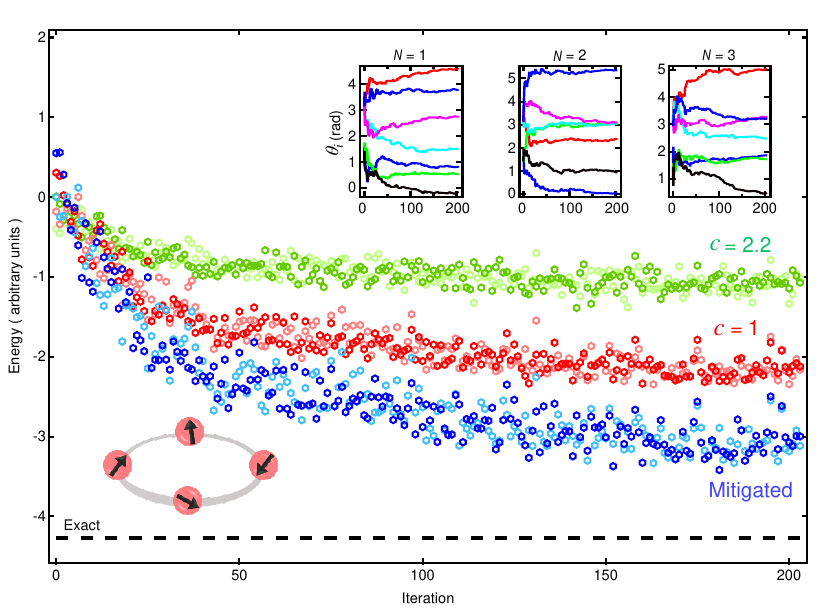}
\caption{Error-mitigated variational optimization of a 4-spin circular Ising model (shown in the bottom left inset) with a transverse field intensity of $J=0.5$. Energy minimization is performed using a 3-layer parameterized circuit tailored to the system. The top insets show the optimization of the variational parameters during the learning process for each layer. At each iteration, the energies of the trial states, measured under different noise coefficients $c=1$ (red) and $c=2.2$ (green), are averaged over 20,000 samples. The mitigated energies (blue), derived via first-order Richardson extrapolation, are fed back into the optimization process to adjust the circuit parameters for subsequent iterations. The darker and lighter data points represent perturbations in opposite directions within the parameter space throughout the optimization process.}
\label{fig:learning}
\end{figure*}

\section{Results}

In our experiments, the dominant source of noise is the longitudinal relaxation of the transmon qubit. The storage cavity has a longitudinal relaxation time that is an order of magnitude longer than that of the transmon qubit, while its transverse relaxation is mainly induced by the transmon qubit due to their dispersive coupling. Given that $T_2^*>T_1$ for the transmon qubit, the longitudinal relaxation of the transmon is the dominant factor. 
Therefore, we focus on mitigating the excited state damping noise $\Gamma_{1}$ of the transmon to suppress the leading imperfections. 

To address this, we implement a noise injection method to the transmon qubit by directly controlling its damping rate through the introduction of an additional population relaxation channel. By leveraging the low-quality factor readout cavity~\cite{Magnard2018PRL_reset,EggerPhysRevApplied2018}, as shown in Fig.~\ref{fig:analogEM}(a), we apply a weak drive with an amplitude $\Omega_\mathrm{ef}$ that is resonant with the transition between the $\ket{e,0}_\mathrm{qr}$ state and the $\ket{f,0}_\mathrm{qr}$ state, facilitating the transfer of population from the first excited state of the transmon to the second excited state $\ket{f}$. Here, $\ket{e,0}_\mathrm{qr}$ represents the joint quantum state of the composite transmon qubit-readout cavity system, as indicated by the subscripts. Subsequently, relaxation from the second excited state of the transmon to the ground state is realized by inducing a decay from the $\ket{f,0}_\mathrm{qr}$ state to $\ket{g,1}_\mathrm{qr}$ through a strong drive with an amplitude $\Omega_\mathrm{f0g1}$ applied to the transition between them. Benefiting from the Purcell effect, the readout cavity has a relatively large decay rate $\Gamma_\mathrm{r}$, and the induced relaxation channel is approximately a Markovian process when the condition $\Omega_\mathrm{ef}\ll\Omega_\mathrm{f0g1}\ll\Gamma_\mathrm{r}$ is satisfied. Therefore, this setup effectively introduces additional damping noise to the computational basis states $\ket{g}$ and $\ket{e}$ of the transmon qubit, with the added damping rate given by $\left(\frac{\Omega_\mathrm{ef}}{\Omega_\mathrm{f0g1}}\right)^2\Gamma_\mathrm{r}$~\cite{SM}. 


However, due to the dispersive coupling, the photon number in the storage cavity influences the transition frequencies in Fig.~\ref{fig:analogEM}(a). Consequently, during the experiment, we extend the gate duration to minimize the photon number in the storage cavity as much as possible, which inevitably increases the impact of decoherence and reduces the gate fidelity~\cite{SM}. 
With the amplitude of $\Omega_\mathrm{f0g1}$ fixed, $T_1$ and $T_2^*$ of the transmon qubit depend on $\Omega_\mathrm{ef}$ for different Fock states in the storage cavity, as illustrated in Figs.~\ref{fig:analogEM}(b) and \ref{fig:analogEM}(c), respectively. We now turn to the evaluation of gate fidelity, as shown in Fig.~\ref{fig:analogEM}(d). \red{It has been shown that in 3D superconducting cavity QED systems,  gate performance is primarily limited by incoherent errors arising from the decoherence of the transmon qubit~\cite{xu2020demonstrationPRL,ma2020errorPASS}. In particular, when the coherence time of the transmon qubit is reduced by a factor of $1/c$, the gate fidelity is expected to degrade by the same factor.} The changes in identity gate fidelity align with the reduction in qubit coherence times as $\Omega_\mathrm{ef}$ varies. However, a notable reduction in Hadamard gate fidelity occurs specifically upon activating the pump $\Omega_\mathrm{f0g1}$. This reduction could be attributed to the potential transitions between the transmon qubit and nearby two-level systems, possibly triggered by the strong multiplexing pump. Despite this issue, we utilize a parameter of $c = 2.2$, where both gate fidelity and coherence time degrade by the same factor, allowing us to proceed with error mitigation.
 
\begin{figure*}
\includegraphics[scale=1]{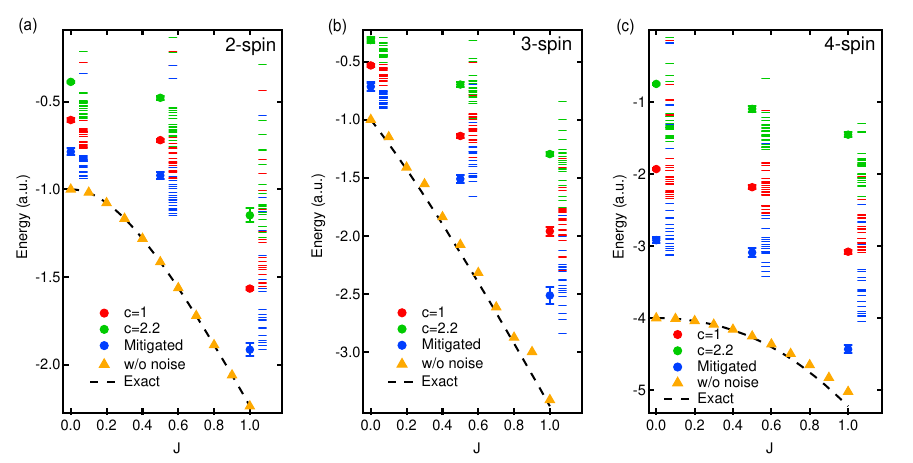}
\caption{Experiment results for the ground state energy of the $2$-spin (a), $3$-spin (b), and $4$-spin (c) circular Ising model with varying transverse field intensities $J$. 
Noise-free simulation results are indicated by the yellow triangles, while the exact ground state energy is shown by the black dashed line. Error-mitigated results obtained through first-order Richardson extrapolation are shown in blue, along with data for noise scaling factors $c=1$ in red and $c=2.2$ in green. \red{The short transverse lines represent numerical simulations incorporating measured decoherence rates, where each data point corresponds to 20 different random initial parameter sets for direct comparison with experimental results.}}
\label{fig:ExpResult}
\end{figure*}
	
We now apply the qubit-efficient VQE approach combined with the analog error mitigation technique to determine the ground state energy of an $N+1$-spin circular Ising model with a transverse field. \red{In this case, the entanglement is typically low except near the quantum critical region. Therefore, outside of this region, it can be expected that a two-qubit system is sufficient to capture the essential features of the system, effectively representing the key dynamics of the model with minimal computational resources.} The Hamiltonian is given by	
\begin{equation}
	H=\sum_{\langle ij\rangle}Z_iZ_j+J\sum_{i}X_i,
	\label{eq:IsingH}
\end{equation} 
where $X$ and $Z$ are Pauli operators, and $J$ is the amplitude of the transverse field. The spin-spin interaction $Z_iZ_j$ is summed over nearest-neighbor pairs $\langle ij\rangle$ on a ring formed by the $N+1$ spins. The quantum circuit, shown in Fig.~\ref{fig:concept}(b), serves as the ansatz for the VQE algorithm, enabling us to generate a parameterized quantum state that approximates the ground state of the system. 

To optimize the circuit parameters, we employ the simultaneous perturbation stochastic approximation algorithm~\cite{Spall1992,kandala2017hardware-efficient}, which is particularly effective in quantum optimization tasks due to its ability to estimate the gradient with only two measurements per iteration, regardless of the number of parameters involved. The red dots in Fig.~\ref{fig:learning} illustrate the learning process of the $4$-spin model, where each iteration of the optimization algorithm includes two measurements corresponding to perturbations in opposite directions in the parameter space, represented by the lighter and darker data points, respectively. Throughout the optimization process, each data point represents the average of more than $20,000$ measurement samples. Given the long lifetime of the photonic qubit in the storage cavity, the standard initialization method, which relies on waiting for spontaneous relaxation of the population in the cavity, would introduce significant delays and potential system drift before convergence. To address this, we implement an active population purge by applying measurement and feedback control to the cavity, allowing for faster initialization~\cite{Pfaff2017,SM}. 

To improve the precision of the VQE, we implement the analog error mitigation that involves measuring the energy of the system under different noise levels and using Richardson extrapolation to estimate the zero-noise energy. Specifically, as previously mentioned, in each iteration, the energy of the quantum state is first obtained under normal noise conditions with a noise coefficient $c=1$, as shown by red dots in Fig.~\ref{fig:learning}. Then, the transmon qubit noise is amplified by a factor of $c=2.2$ through the noise injection, and the energy is recalculated. 
Using Richardson extrapolation, we estimate the zero-noise energy by combining these two measurements, mitigating the impact of transmon qubit noise to the first order. The mitigated energies are then iteratively incorporated into the optimization process to refine the circuit parameters, leading to improved accuracy of the results. Despite these efforts, a discrepancy remains compared to the exact energy, primarily due to the extended gate duration, which increases the impact of decoherence. While error mitigation reduces noise effects, first-order mitigation alone is insufficient for achieving the desired precision, underscoring the need for higher-order techniques to close the gap. 



We further validate the qubit-efficient VQE algorithm and explore the effect of analog error mitigation by solving the ground state energy of systems with varying spin numbers ($N=1,2,3$) and different transverse field intensities $J$. Figure~\ref{fig:ExpResult} summarizes the experimental results, demonstrating the effectiveness of our approach across a range of system sizes and parameters. For each system configuration, we perform numerical simulations incorporating the measured system decoherence rates, using 20 random initial parameter sets. The corresponding simulation data (horizontal bars) are plotted to the right of the experimental data points for direct comparison. 

As evident from Fig.~\ref{fig:ExpResult}, the experimental data exhibit a consistent trend with the exact ground state energy as the transverse field intensity $J$ varies. However, a systematic deviation from the ideal eigenvalues is observed, which can be attributed to the presence of unmitigated noise in the quantum system. Remarkably, our numerical simulations, which take into account the system's decoherence rates, show excellent agreement with the experimental data. This consistency highlights the accuracy of our modeling and the predominant role of decoherence in limiting the performance of the VQE algorithm. For comparison, we also perform numerical simulations of the circuit using an ideal noise-free model, as shown by the yellow triangles in Fig.~\ref{fig:ExpResult}. The close alignment between the exact ground state energies and the simulation with ideal model confirms the effective of the qubit-efficient VQE algorithm.

\red{In evaluating the scalability of our approach, we note that systems with bounded entanglement, such as 1D cluster states~\cite{LiuPhysRevResearch2019}, allow the MPS-based representation to scale effectively with minimal loss of accuracy. Since the transverse field Ising model exhibits low entanglement away from the quantum critical region, our circuit is expected to efficiently simulate larger spin systems without substantial accuracy loss.} To assess the potential for further enhancing the accuracy of our qubit-efficient VQE, we perform numerical simulations incorporating higher-order error mitigation techniques~\cite{SM}. These simulations indicate that extending the analog error mitigation to higher orders can significantly reduce the remaining discrepancies, bringing the mitigated results closer to the ideal ground state energies. While experimental implementation of higher-order mitigation is not feasible in the current setup due to practical limitations, these simulation results highlight promising avenues for future research in error mitigation techniques.

\section{Conclusion}

In this work, we experimentally demonstrated a qubit-efficient VQE algorithm on a superconducting quantum processor, addressing the challenges posed by the limited qubit resources and high noise levels of NISQ devices. By leveraging an MPS representation, we efficiently simulated the ground state energies of an $N+1$-spin circular transverse-field Ising model using only two physical qubits: a transmon qubit and a high-coherence photonic mode. We validated the qubit-efficient VQE and analog error mitigation approach by determining the ground state energies of the 4-spin Ising model, demonstrating improved precision compared to unmitigated results. 

The methods developed here can be readily extended to other quantum algorithms and larger system sizes~\cite{GuoNP2024,Au-Yeung2024}, paving the way for tackling more complex problems in quantum chemistry, condensed matter physics, and optimization. However, it is important to recognize the limitations of the current approach. The reliance on the MPS representation restricts the class of quantum states that can be efficiently encoded, particularly those with high entanglement across all qubits. Additionally, while the analog error mitigation protocol is effective for suppressing certain types of noise, it may not be sufficient for mitigating all sources of errors in the quantum system. Overcoming these limitations will require the development of more advanced qubit encodings, enhanced error mitigation techniques, and the potential integration with quantum error correction schemes. 


	

\begin{acknowledgments}
This work was funded by the National Natural Science Foundation of China (Grants No. 11925404, 92165209, 92365301, 92265210, 11890704, 92365206, 12474498, T2225018, 92270107, 12188101, T2121001, 62173201), Innovation Program for Quantum Science and Technology (Grant No.~2021ZD0300200 and 2021ZD0301800), and the National Key R\&D Program (2017YFA0304303). This work was also supported by the Fundamental Research Funds for the Central Universities and USTC Research Funds of the Double First-Class Initiative. This work was partially carried out at the USTC Center for Micro and Nanoscale Research and Fabrication.
\end{acknowledgments}


%

\cleardoublepage
\onecolumngrid
\begin{center} 
\textbf{\Large{}Supplementary Information for ``Experimental Implementation of a Qubit-Efficient Variational Quantum Eigensolver with Analog Error Mitigation on a Superconducting Quantum Processor''}{\Large\par}
\par\end{center}

\vspace{7em} 
\twocolumngrid

\section{Controlled damping technique}

\begin{figure} [b]
	\includegraphics{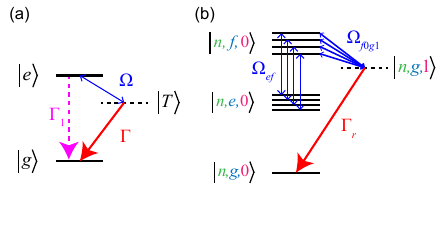}
	\caption{(a) Scheme of the damping enhancement of a two-level qubit via a transient level $\ket{T}$. (b) Scheme of the damping enhancement in the experiment. \red{$\ket{s,q,r}$ represents a product state, where $s$ denotes the excitation in the storage cavity, $q$ denotes the transmon qubit state, and $r$ denotes the excitation in the readout cavity.}}
	\label{fig:supple1principle}
\end{figure}
In our experiment, we control the damping rate $\Gamma_1$ of the transmon qubit by adjusting its coupling strength to the low-Q readout cavity~\cite{pechal2014microwave, Magnard2018PRL_reset, EggerPhysRevApplied2018,sunada2022fast,marques2023all}. Figure~\ref{fig:supple1principle}(a) illustrates the basic scheme of the controlled damping process. The qubit is described by two energy levels, $\ket{g}$ and $\ket{e}$, along with a transient level $\ket{T}$, which decays rapidly to the ground state $\ket{g}$ with a large decay rate $\Gamma$. The transition between the excited state $\ket{e}$ and $\ket{T}$ is driven with an amplitude $\Omega$. Then the system evolution is governed by
\begin{equation}
\dot{\rho}=-i\left[\Omega\left(\ket{e}\bra{T}+\ket{T}\bra{e}\right),\rho\right]+\Gamma\mathcal{L}\{\ket{g}\bra{T}\}(\rho),
\label{eq:modelEvolution}
\end{equation}
where $\mathcal{L}$ is the Lindblad operator representing the decay process. 
For a sufficiently small transition amplitude $\Omega\ll\Gamma$, the population in the transient state $\ket{T}$ remains low. Introducing a small parameter $\delta$, we can expand the expression of the density matrix of the system $\rho$ as 
\begin{align}
\rho=&\rho_\mathrm{gg}\ket{g}\bra{g}+\rho_\mathrm{ge}\ket{g}\bra{e}+\rho_\mathrm{eg}\ket{e}\bra{g}+\rho_\mathrm{ee}\ket{e}\bra{e} \nonumber \\
&+\delta\left(\rho_\mathrm{gT}\ket{g}\bra{T}+\rho_\mathrm{Tg}\ket{T}\bra{g}+\rho_\mathrm{eT}\ket{e}\bra{T}+\rho_\mathrm{Te}\ket{T}\bra{e}\right) \nonumber \\
&+\delta^2\rho_\mathrm{TT}\ket{T}\bra{T}.
\end{align}
Substituting this expanded form into Eq.~\ref{eq:modelEvolution}, we obtain the following set of differential equations for each matrix element
\begin{align}
	&\dot{\rho}_\mathrm{gg}=\Gamma\delta^2\rho_\mathrm{TT}, \nonumber \\
	&\dot{\rho}_\mathrm{ee}=-i\Omega\delta\left(\rho_\mathrm{Te}-\rho_\mathrm{eT}\right), \nonumber \\
	&\dot{\rho}_\mathrm{ge}=i\Omega\delta\rho_\mathrm{gT}, \nonumber \\
	&\delta\dot{\rho}_\mathrm{gT}=i\Omega\rho_\mathrm{ge}-\frac{\Gamma\delta}{2}\rho_\mathrm{gT}, \nonumber \\
	&\delta\dot{\rho}_\mathrm{eT}=-i\Omega\delta^2\rho_\mathrm{TT}+i\Omega\rho_\mathrm{ee}-\frac{\Gamma\delta}{2}\rho_\mathrm{eT}, \nonumber \\
	&\delta^2\dot{\rho}_\mathrm{TT}=i\Omega\delta\left(\rho_\mathrm{Te}-\rho_\mathrm{eT}\right)-\Gamma\delta^2\rho_\mathrm{TT}. 
\end{align}
By adiabatically eliminating the transient level and assuming the matrix elements related to $\ket{T}$ reach a steady state quickly  $(\dot{\rho}_\mathrm{gT},\,\dot{\rho}_\mathrm{eT},\,\dot{\rho}_\mathrm{TT}=0)$, we derive the evolution of the qubit states as
\begin{align}
	&\dot{\rho}_\mathrm{gg}=-\dot{\rho}_\mathrm{ee}=\frac{4\Omega^2\Gamma}{4\Omega^2+\Gamma^2}\rho_\mathrm{ee}\approx\frac{4\Omega^2}{\Gamma}\rho_\mathrm{ee}, \nonumber \\
	&\dot{\rho}_\mathrm{ge}=-\frac{2\Omega^2}{\Gamma}\rho_\mathrm{ge}.
\end{align}
These results describe the effective dynamics within the qubit subspace under controlled damping, allowing us to adjust the relaxation rate of the qubit \red{$\Gamma_1=\frac{4\Omega^2}{\Gamma}$} by tuning the coupling and drive parameters.

The controlled damping technique used in our experiment is shown in  Fig.~\ref{fig:supple1principle}(b) and can be understood through the simplified model described earlier in Fig.~\ref{fig:supple1principle}(a). First, we increase the damping rate of \red{the second excited state of the transmon qubit} $\ket{f}_\mathrm{q}$ by coupling it to the readout cavity using a microwave pump with an amplitude $\Omega_\mathrm{f0g1}$. \red{This coupling modifies $\ket{f}_\mathrm{q}$ to exhibit a damping rate of $4\Omega_\mathrm{f0g1}^2/\Gamma_\mathrm{r}$ through the transient level $\ket{ng1}_\mathrm{sqr}$, where $\Gamma_\mathrm{r}$ is the damping rate of the readout cavity}. Next, the damping rate of the $\ket{e}_\mathrm{q}$ state is increased by coupling it to $\ket{f}_\mathrm{q}$ with an amplitude $\Omega_\mathrm{ef}$, leading to a damping rate for $\ket{e}_\mathrm{q}$ of $\frac{\Omega_\mathrm{ef}^2}{\Omega_\mathrm{f0g1}^2}\Gamma_\mathrm{r}$.

We also account for the photon-number-dependent frequency shift of the transmon qubit due to its dispersive coupling to the storage cavity. Each photon added to the storage cavity shifts the frequency of the transmon qubit by $\chi_{\mathrm{qs}}$. This effect requires the microwave pump to operate at different frequencies depending on the Fock states in the storage cavity~\cite{gertler2021protecting}. Since the computational basis used in this experiment consists of Fock states $\ket{0}_\mathrm{s}$ and $\ket{1}_\mathrm{s}$, at least two distinct pump frequencies are necessary. In addition, the Hadamard gate in this experiment is implemented using the gradient ascent pulse engineering method (GRAPE)~\cite{Khaneja2005,deFouquieres2011}, during which the storage cavity may contain a higher photon number. To ensure the controlled damping remains effective during the GRAPE pulse, the microwave pump must cover the frequency range corresponding to the largest photon number present during the evolution. To reduce photon occupation, we extend the GRAPE pulse duration, which results in a lower gate fidelity, and ultimately settle on a pulse length of $3~\mu$s after balancing these factors. The numerical simulations show that the photon number remains below four during the evolution, with the population of the $\ket{4,e}_\mathrm{sq}$ state being less than $2\%$. 

\begin{figure}
	\includegraphics{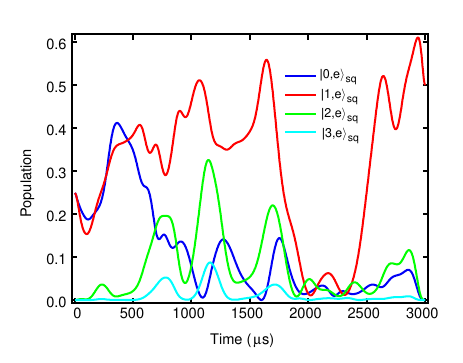}
	\caption{Population dynamics 
 of the $\ket{n,e}_\mathrm{sq}$ states for $n=0,1,2,3$ during the GRAPE pulse for the Hadamard gate, with the initial state being $(\ket{g}+i\ket{e})(\ket{0}-\ket{1})/2$. The $\ket{3,e}_\mathrm{sq}$ state is observed to have the highest population fraction of $8.73\%$ among all 36 initial cardinal states. The populations for states with larger photon numbers are smaller, with the maximum population of the $\ket{4,e}_\mathrm{sq}$ state being only $1.56\%$.}
	\label{fig:supple3GRAPEpopulation}
\end{figure}

\section{Experimental setup}

This experiment is implemented on a sample based on a 3D superconducting \red{cavity-QED architecture~\cite{ma2020errorPASS,BlaisCQEDRMP}}, in which a transmon qubit is dispersively coupled to a high-Q 3D storage cavity and a low-Q strip-line readout cavity. As described in the main text, the transmon qubit serves as the reusable qubit and the lowest two Fock states of the storage cavity form the mediating qubit. The device parameters and coherence times are presented in Table~\ref{Table:DeviceParameters} and Table~\ref{Table:coherenttime}. Since the transmon qubit experiences significantly greater decoherence than the storage cavity, the experiment focuses on mitigating errors from the transmon qubit by controlling its decoherence and using noise-extrapolation techniques to counterbalance the deleterious effects.

\begin{table}
	\centering %
	\begin{tabular}{cccccc}
		\hline 
		Term  & Measured &  &  &  & \tabularnewline
		\hline 
		qubit frequency $\omega_{\mathrm{q}}/2\pi$ & 5.801 GHz &  &  &  & \tabularnewline
		storage cavity frequency $\omega_{\mathrm{s}}/2\pi$ & 6.571 GHz &  &  &  & \tabularnewline
		readout cavity frequency $\omega_{\mathrm{r}}/2\pi$ & 8.9097 GHz &  &  &  & \tabularnewline
		\hline 
		self-Kerr of the qubit $K_{\mathrm{q}}/2\pi$ & 241 MHz &  &  &  & \tabularnewline
		self-Kerr of the storage cavity $K_{\mathrm{s}}/2\pi$ & 2.9 kHz &  &  &  & \tabularnewline
		\hline 
		cross-Kerr between the qubit\\ and the storage cavity $\chi_{\mathrm{qs}}/2\pi$ & 0.945 MHz &  &  &  & \tabularnewline
		cross-Kerr between the qubit\\ and the readout cavity $\chi_{\mathrm{qr}}/2\pi$ & 1.3 MHz &  &  &  & \tabularnewline
		\hline 
	\end{tabular}\caption{Device parameters.}
	\label{Table:DeviceParameters} 
\end{table}
\begin{table}
	\centering %
	\begin{tabular}{cccccc}
		\hline 
		$\quad$ & $Q$ & $S$ & $R$ &  & \tabularnewline
		\hline 
		$T_{1}$ & $24\ \mu$s & $740\ \mu$s & $68$ ns &  & \tabularnewline
		$T_{2}^*$ & $28\ \mu$s & $510\ \mu$s & - &  & \tabularnewline
		\hline 
		thermal excitation & $3.4\%$ & $<1\%$ & - &  & \tabularnewline
		\hline 
	\end{tabular}\caption{Coherence time and thermal population of the transmon qubit ($Q$), the storage cavity ($S$), and the readout cavity ($R$).}
	\label{Table:coherenttime} 
\end{table}

\begin{figure} [b]
	\includegraphics{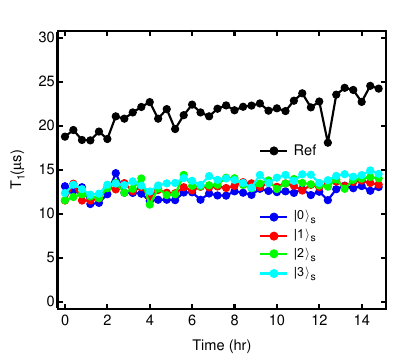}
	\caption{Tracking $T_1$ of the transmon qubit for different Fock states in the storage cavity.}
	\label{fig:supple2dampenhance}
\end{figure}

\begin{figure*}
	\includegraphics{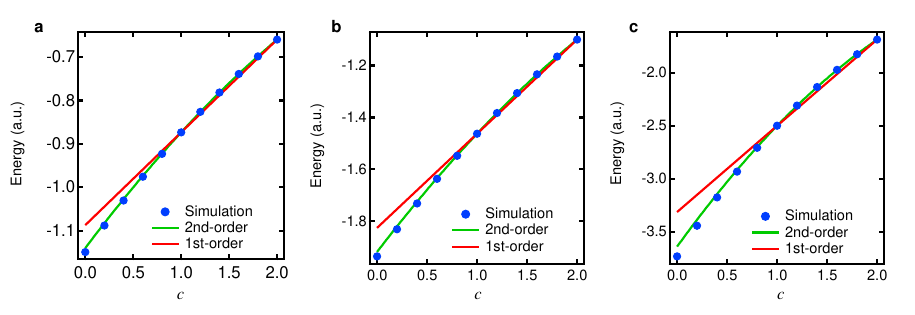}
	\caption{Simulation results for the ground-state energy across varying noise amplification factors $c$, with transverse field intensity fixed at $J=0.5$. Panel (a) corresponds to a 2-spin system, (b) to a 3-spin system, and (c) to a 4-spin system. The blue dots represent the optimized ground-state energy for $c$ values ranging from 0.0 to 2.0. The first-order extrapolations (red solid lines) are based on the data points for $c=1.0$ and $c=2.0$. The second-order extrapolations (green solid lines) utilize the data points for $c$ values including 1.0, 1.2, ..., up to 2.0.}
	\label{fig:supple5_1order_extrapolation}
\end{figure*}

In our experiment, the high-power microwave pumps for $\Omega_\mathrm{f0g1}$ are realized using four independent microwave drives, each with equal amplitude and frequency intervals. The low-power microwave pumps for $\Omega_\mathrm{ef}$ consist of four sidebands generated from a single microwave drive, realized using by two AWG channels and one IQ mixer, with equal amplitude and frequency differences between the sidebands.  
\red{The power and frequency interval of the $\Omega_\mathrm{f0g1}$ are calibrated to achieve the maximum the damping effect on the state $\ket{n,f}_\mathrm{sq}$. After calibration, the damping time  $T_{\ket{1,f}}$ is approximately $2~\mu$s.} In the experiment, we maintain a fixed amplitude of $\Omega_\mathrm{f0g1}$ while controlling the damping rate of the transmon qubit through the amplitude of $\Omega_\mathrm{ef}$, as shown in the main text.

\red{Fluctuations in coherence and relaxation times, which are common in superconducting qubit systems, can affect the accuracy of quantum operations and the stability of results over time.} We monitor the coherence of the transmon qubit over time.  Figure~\ref{fig:supple2dampenhance} tracks $T_1$ of the transmon qubit without the controlled damping as a reference, comparing it to the reduced $T_1$ values for each Fock state. Over a period of 16 hours, the system exhibits small drift, with the intrinsic $T_1$ increasing from $20~\mu$s to $25~\mu$s, while the reduced $T_1$ also shows a slight increase. Although the ratio between them does not remain perfectly constant, the method demonstrates reasonable robustness. \red{Furthermore, in this experiment, we implement a fast initialization method on the storage cavity to improve the reliability of error mitigation and reduce susceptibility to coherence variations.}

\section{Numerical simulation}

We perform numerical simulations that replicate the experimental sequence using experimentally calibrated Hamiltonian parameters. 
We iterate the entire process 200 times using \red{the simultaneous perturbation stochastic approximation algorithm~\cite{Spall1992,kandala2017hardware-efficient}} and compare the results with our experimental data. The optimized ground state energies are calculated for systems with varying spin numbers ($N=2,3,4$) at a fixed transverse field intensity of $J=0.5$, considering different noise amplification factors $c$, as shown in Fig.~\ref{fig:supple5_1order_extrapolation}.

A second-order polynomial fit is conducted on the simulated data across noise amplification factors $c$ ranging from 1.0 to 2.0, yielding an extrapolated value that closely approximates the simulated zero-noise energy. In accordance with the experimental procedure, a first-order extrapolation is also performed on the simulated data. The discrepancy observed between the first-order extrapolated results and the simulated zero-noise energy highlights the impact of higher-order noise effects. This observation underscores the need for higher-order extrapolation methods to achieve more accurate noise mitigation in quantum simulations.





\end{document}